\documentclass[11pt,twoside]{article}
\usepackage{./asp2014}

\aspSuppressVolSlug
\resetcounters

\begin{document}

\title{Atlas of the     solar intensity spectrum and its center to limb variation}
\author{R. Ramelli,$^1$ M. Setzer,$^{1,2}$ M. Engelhard,$^2$ M.Bianda,$^1$ F.~Paglia,$^1$ J.~O.~Stenflo,$^{1,3}$, G.~K\"uveler,$^2$ R. Plewe$^2$}

\affil{$^1$Istituto Ricerche Solari Locarno IRSOL, associated to Universit\`a della Svizzera
  Italiana, Locarno, TI, Switzerland;
}
\affil{$^2$Hochschule RheinMain, R\"usselsheim, Deutschland;
}
\affil{$^3$Institute of Astronomy, ETH Zurich, 8093 Zurich, Switzerland}

\paperauthor{Renzo Ramelli}{ramelli@irsol.ch}{0000-0002-1976-1024}{Universit\`a della Svizzera Italiana}{Istituto Ricerche Solari Locarno, IRSOL}{Locarno}{TI}{6605}{Switzerland}
\paperauthor{Michele Bianda}{mbianda@irsol.ch}{}{Universit\`a della Svizzera Italiana}{Istituto Ricerche Solari Locarno, IRSOL}{Locarno}{TI}{6605}{Switzerland}
\paperauthor{Martin Setzer}{setzermartin@googlemail.com}{}{Hochschule
  RheinMain}{Fachbereich Ingenieurwissenschaften}{R\"usselsheim}{}{65428}{Deutschland}
\paperauthor{Mathis Enegelhard}{mathis.engelhard@web.de}{}{Hochschule
  RheinMain}{Fachbereich Ingenieurwissenschaften}{R\"usselsheim}{}{65428}{Deutschland}
\paperauthor{Filippo Paglia}{filippo.paglia@gmail.com}{}{Universit\`a della
  Svizzera Italiana}{Istituto Ricerche Solari Locarno,
  IRSOL}{Locarno}{TI}{6605}{Switzerland}
\paperauthor{Jan Olof Stenflo}{stenflo@irsol.ch}{0000-0002-5524-6628}{Universit\`a della Svizzera Italiana}{Istituto Ricerche Solari Locarno, IRSOL}{Locarno}{TI}{6605}{Switzerland}
\paperauthor{Gerd K\"uveler}{kueveler@googlemail.com}{}{Hochschule
  RheinMain}{Fachbereich Ingenieurwissenschaften}{R\"usselsheim}{}{65428}{Deutschland}
\paperauthor{Rouven Plewe}{}{}{Hochschule
  RheinMain}{Fachbereich Ingenieurwissenschaften}{R\"usselsheim}{}{65428}{Deutschland}

\begin{abstract}
The solar limb darkening function is well
known and is widely employed in models of the solar atmosphere. 
However, there has been a lack of systematic spectrally resolved measurements.
Therefore we recently decided to start an observing campaign 
with the Gregory Coud\'e Telescope at IRSOL in Locarno in order to
produce a spectral atlas
obtained at 10 different heliocentric angles $\theta$, chosen so that 
$\mu = \cos\theta$  covers the interval from 0.1 to 1.0 in step of 0.1. 
The measurements
carried out till now include the spectral range from 439 nm to 666 nm.
 
The collected data provide information about the
anisotropy of the emergent radiation field on the solar surface, allowing a
better modeling of the Second Solar Spectrum. In addition the data
give observational constraints that should be taken into account when 
modeling the solar atmosphere.
\end{abstract}

\section{Introduction}

The Sun's disk observed at visible wavelengths is limb darkened. The main
reason is the temperature decrease with height in the layers from where the
radiation is coming, when they are seen at different center-to-limb positions,
which are usually defined by $\mu = \cos\theta$, where $\theta$ is
the heliocentric angle. Due to the fact that the opacity, the
height of formation and non-local thermodynamical equilibrium (non-LTE)
effects vary along the profiles of the different lines present in
the solar spectrum, the limb darkening function changes remarkably across the
spectral lines. 

The limb darkening is also directly related to the anisotropy of the emergent
radiation at the solar surface, which is the source of the scattering
polarization whose spectral dependence is referred to as the Second Solar
Spectrum \citep{stenflo97}. Spectrally resolved limb darkening measurements 
thus provide important information to be taken into
account when modeling both the Second Solar Spectrum and the solar atmosphere.

The center-to-limb variation (CLV) of the intensity solar
spectrum is richly structured, 
but in ways that differ profoundly from the usual intensity
spectrum (in this context, the First Solar Spectrum) and from the Second Solar
Spectrum. Thus
\citet{Stenflo15} suggests referring to the intensity CLV as the Third
Solar Spectrum (SS3).

\section{Observations}

The SS3 atlas presented in this work has been obtained with the Gregory-Coud\'e telescope at IRSOL in
the spectral range from 439 nm to 666 nm
at 10 different heliocentric angles $\theta$ corresponding to $\mu$ from 0.1 to 1.0
in step of 0.1. The 
spectrograph connected to the telescope is a Czerny Turner of  10 m focal length equipped with a
grating with 316 grooves/mm, a blaze angle of 63$^\circ$ and whose size is 
180 x 360 mm. The observations have been recorded
with a ZIMPOL camera \citep{zimpol-spie2010} that usually is used with masked
CCD sensors for polarimetry, but which for this project has been equipped with
an unmasked sensor that is better suited for intensity measurements. The data
acquisition procedure has been almost fully automatized thanks to the usage of
the scripting possibilities given by the ZIMPOL control software described by
\citet{zimpol-spie2010}. Accurate positioning has been done thanks to the
 Primary Image Guiding system reported by \citet{PIG11}.
A limb tracking system connected to the slit-jaw camera and 
based on a tilting glass plate, has been used  to keep a constant limb distance,
when the limb was visible in the field of view of the telescope
(i.e. from $\mu = 0.1$ to $\mu = 0.4$).
A Dove prism has been used to rotate the image observed in order to keep the
orientation of the heliographic North parallel to the spectrograph slit.
 The observing sequence was chosen in order to  
have an observation at the center of the solar disc ($\mu=1$),
immediately before or after each
observation taken at the different $\mu$-positions. For each spectral window
we also took a corresponding flat-field and dark observation. After a quality
check some observations needed to be repeated.

\section{Data reduction and results}

Each observation has been corrected for flat-field and dark.
The spectral intensity profiles have been obtained averaging over 30
arcseconds along the spatial direction. Both the spectra $I_\mu(\lambda)$ at
the different $\mu$ positions and the spectra $I_c(\lambda)$ at the disc center
have been normalized so that the continuum is set to 1. The wavelength
scale has been determined with the help of the FTS atlas \citep{Kurucz}.
For each $\mu$-position we calculate the ratio

\begin{equation}
R_\mu(\lambda) = \frac{I_\mu(\lambda)}{I_c(\lambda)}
\end{equation}

\begin{figure}[htb]
\begin{center}
\includegraphics[width=0.9\linewidth,height=13.cm]{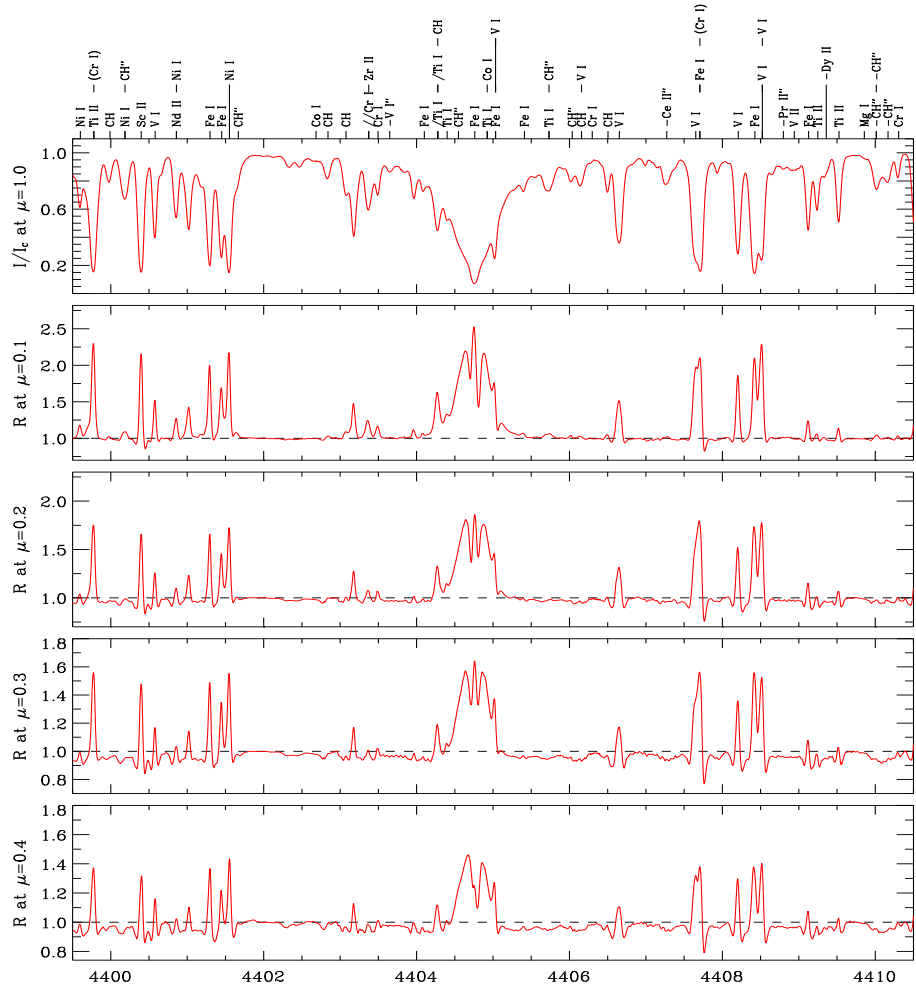}
\caption{\label{fig1} Measurement example of SS3 atlas. Top panel: spectrum observed at disc center. Other
  panels: $R_\mu(\lambda)$ for $\mu=0.1;0.2;0.3;0.4$}
\end{center}
\end{figure}

\begin{figure}[htb]
\begin{center}
\includegraphics[width=0.9\linewidth,height=13.4cm]{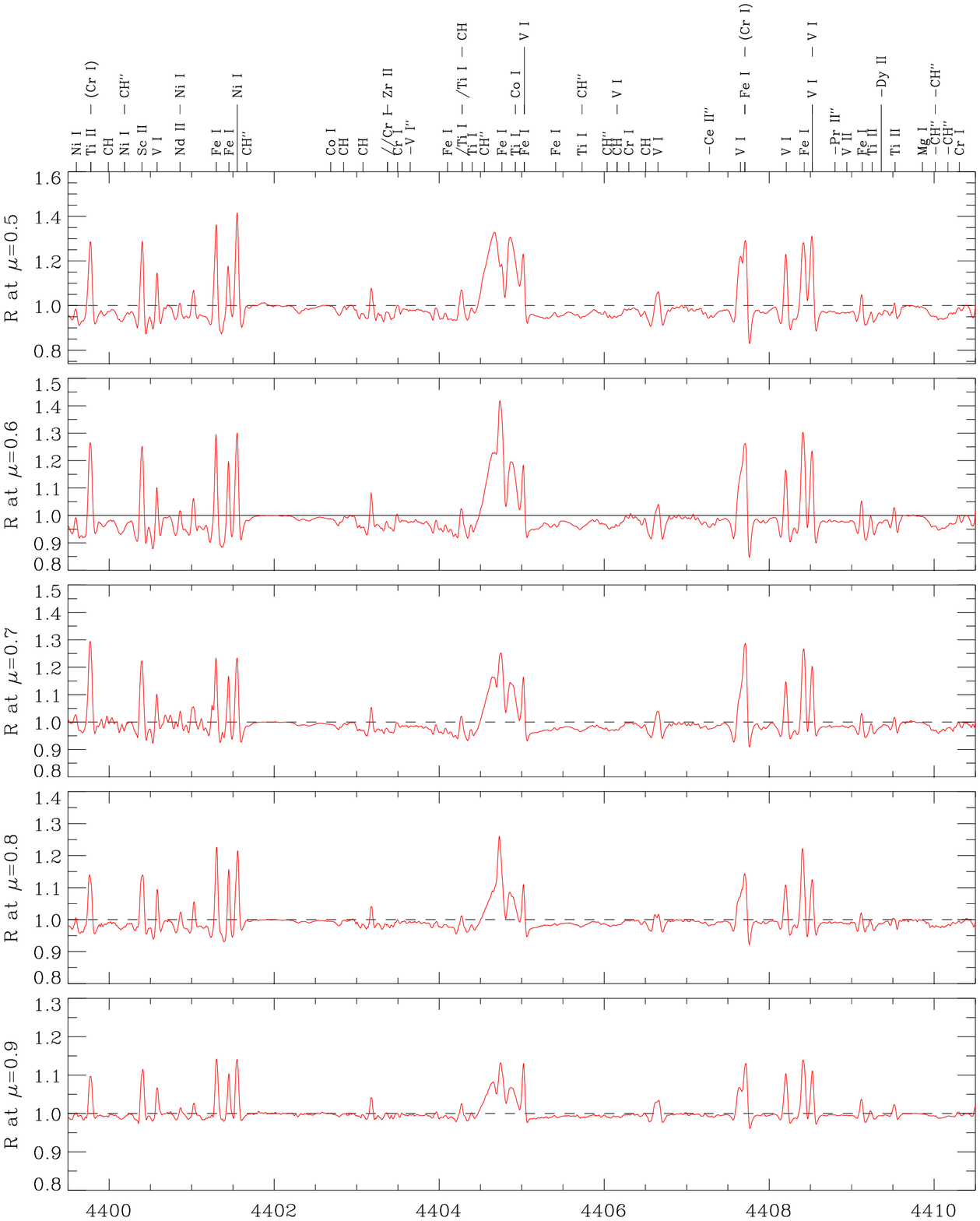}
\caption{\label{fig2} Measurement example of SS3 atlas: $R_\mu(\lambda)$ for $\mu=0.5;0.6;0.7;0.8;0.9$}
\end{center}
\end{figure}

The SS3 atlas obtained in this work displays the intensity spectrum at disc
center $I_c(\lambda)$ and the 9 $R_\mu$-profiles obtained at the different
center to limb positions (see example in Figures \ref{fig1} and \ref{fig2}).
Our effort has been put in measuring precisely as possible
the $R_\mu(\lambda)$ function. Since very precise and reliable 
measurements have been made by \citet{Neckel}
for center to limb variation 
in the continuum $I^N_\lambda(\mu)$ and by \citet{Kurucz}
for the intensity spectrum at disc center $I^{FTS}_c(\lambda)$,
 we suggest the reader that he should use 
for the intensity spectrum at a particular $\mu$ position
\begin{equation}
I_\mu(\lambda) = R_\mu(\lambda) \cdot I^N_\lambda(\mu) \cdot I^{FTS}_c(\lambda)
\end{equation}
as best estimate.
   
Data are made available to the public on the IRSOL web-site at the address:

\vspace{3mm}
\url{http://www.irsol.ch/data-archive}.\\[3mm]

\acknowledgements 
IRSOL is supported by the Swiss Confederation (SEFRI) , Canton Ticino, the city of Locarno and the local municipalities.
This research work was financed by SNF grants 200020\_157103  and  
200020\_169418. We are grateful to the Fondazione Aldo e Cele Dacc\`o for their
financial contribution.

\newpage

\bibliography{ramelli_2}

\end{document}